\documentclass{jfm}
\usepackage[authoryear,round]{natbib}

\usepackage{graphicx}
\usepackage{amsmath}
\usepackage{color}

\newcommand{\Eqs}{Equations~}
\newcommand{\eq}{equation~}
\newcommand{\eqs}{equations~}

\newcommand{\fig}{figure~}

\makeatletter
\newcommand\abbr{\@ifnextchar.{}{.\@}}
\makeatother
\renewcommand\eg{e.g\abbr}
\newcommand\ie{i.e\abbr}

\newcommand{\PI}{pseudo-incompress\-ible}

\newcommand{\Ray}{\mbox{\textit{Ra}}}
\newcommand{\Tay}{\mbox{\textit{Ta}}}

\newcommand{\dd}{\mathrm{d}}

\newcommand{\ii}{\mathrm{i}}
\newcommand{\kh}{k_{\rm h}}

\newcommand{\cp}{c_p}
\newcommand{\cv}{c_v}
\newcommand{\N}{\mathcal{N}}

\begin{document}

\title
  [Oscillatory convection and limitations of the Boussinesq approximation]
  {Oscillatory convection and limitations of the Boussinesq approximation}
\author
  [T.~S.~Wood and P.~J.~Bushby]
  {\sc T.~S.~Wood and P.~J.~Bushby}
\affiliation{School of Mathematics and Statistics,
Newcastle University, Newcastle upon Tyne, NE1 7RU}
\maketitle

\begin{abstract}
We determine the asymptotic conditions under which the Boussinesq approximation is valid for oscillatory convection in a rapidly rotating fluid. In the astrophysically relevant parameter regime of small Prandtl number, we show that
the Boussinesq prediction for the onset of convection is valid only under much more restrictive conditions than those that are usually assumed.
In the case of an ideal gas,
we recover
the Boussinesq results only if
the ratio of the domain height to a typical scale height
is much smaller than the Prandtl number.
This requires an extremely shallow domain in the astrophysical parameter regime.
Other commonly-used ``sound-proof'' approximations
generally perform no better than the Boussinesq approximation.
The exception is a particular implementation of the \PI\ approximation,
which predicts the correct instability threshold beyond the range of validity of the Boussinesq approximation.

\end{abstract}

\hrule

\section{Introduction}
Most astrophysical objects contain regions in which heat is transported by convection.
The numerical modelling of these convective flows (which are usually turbulent)
is difficult because of the stiffness of the governing equations caused by the presence of acoustic waves.
Almost all convection models therefore filter out these waves by using a ``sound-proof'' set of equations,
such as the Boussinesq, anelastic or \PI\ equations.
Each of these sound-proof approximations is founded on certain physical assumptions,
which may not be valid in all cases of interest.
Specifically, the Boussinesq equations  \citep[\eg][]{SpiegelVeronis60}
are valid only for small perturbations to the thermodynamic variables in systems with small vertical lengthscales
(in particular, the domain height must be much smaller than all of the thermodynamic scale heights).
The anelastic equations require less restrictive assumptions, but do require the fluid to be nearly adiabatically stratified \citep[\eg][]{OguraPhillips62,LippsHemler82}.
The \PI\ equations were introduced by \citet{Durran89} as an improvement to the anelastic equations.
Although they are formally valid only under the same conditions as the anelastic equations,
albeit for stratifications stronger than anticipated by standard asymptotics \citep{Klein-etal10},
in some cases they are found to better approximate the true dynamics \citep[\eg][]{Achatz-etal10}.
This situation is further complicated by the fact that there are several different versions of both the anelastic and \PI\ equations currently in use, with no general consensus on which is ``best''
\citep[\eg][]{Brown-etal12,Vasil-etal13}.

The interiors of stars and gaseous planets
are characterised by rapid rotation and low viscosity.
In this parameter regime, convection is often oscillatory close to onset \citep[\eg][]{Jones-etal09}.
The first studies of oscillatory convection were performed under the Boussinesq approximation
\citep{Chandrasekhar53}, but it has never been determined in precisely what asymptotic limit the Boussinesq
and fully compressible results agree.
Perhaps surprisingly,
some implementations of the anelastic approximation
exhibit unphysical ``negative Rayleigh number'' convection in this parameter regime
\citep{Drew-etal95,Calkins-etal15}.
As other implementations do not
exhibit this unphysical behaviour \citep{Jones-etal09},
it seems that oscillatory convection is an important test case for comparing different sound-proof models.

In this paper, we perform a careful analysis of the onset of oscillatory convection in the fully compressible
system, in order to determine the precise conditions under which the Boussinesq results are valid.
We find that these conditions are much more stringent than anticipated from the usual heuristic arguments.
In particular, it is \emph{not} sufficient that the vertical scale of the domain is much smaller than the
thermodynamic scale heights.
This analysis is then extended to a very general set of sound-proof equations,
which includes the anelastic
and \PI\ approximations as special cases.
In doing so, we introduce a simple but powerful technique
that can be used to standardise the anelastic and \PI\ equations,
building on an observation of \citet{OguraPhillips62} about energy conservation,
thus removing any ambiguity in their formulation.
Our standardised anelastic and \PI\ equations are both free from the unphysical behaviour noted
by \citet{Calkins-etal15}.
However, the standardised \PI\ equations are the only ``sound-proof'' system
that correctly predicts the threshold for oscillatory convection
on larger vertical scales than the Boussinesq approximation.

\section{Fully compressible vs.~Boussinesq}

\subsection{The governing equations}

In order to determine the true convective stability threshold,
we linearise the fully compressible equations about a hydrostatic background
in a reference frame rotating with angular velocity $\boldsymbol{\Omega}$.
For simplicity we neglect self-gravity and the centrifugal force, so the gravitational acceleration $\boldsymbol{g}$
points directly downward.
We adopt Cartesian coordinates in which $\boldsymbol{g} = -g\boldsymbol{e}_z$ and $\boldsymbol{\Omega} = \Omega\boldsymbol{e}_z$,
and write the linearised equations in the form
\begin{align}
  \frac{\partial\boldsymbol{u}}{\partial t} + 2\boldsymbol{\Omega}\times\boldsymbol{u}
  &= - s_1\bnabla T_0 + T_1\bnabla s_0 - \bnabla(p_1/\rho_0)
    + \boldsymbol{f}_1 \label{eq:FC_mom} \\
  \frac{\partial\rho_1}{\partial t} + \bnabla\bcdot(\rho_0\boldsymbol{u}) &= 0 \label{eq:mass} \\
  \frac{\partial s_1}{\partial t} + \boldsymbol{u}\bcdot\bnabla s_0
  &= \frac{q_1}{T_0}, \label{eq:entropy}
\end{align}
where subscripts 0 and 1 refer to unperturbed quantities (which are functions of $z$ only) and their linear perturbations, respectively.
The quantities $\boldsymbol{f}_1$ and $q_1$ are the viscous force and heating rate per unit mass,
whose precise form we specify later.
In order to close these equations we require relations between the thermodynamic perturbations.
For convenience we introduce the specific enthalpy, $H(p,s)$,
which is a function of pressure, $p$, and specific entropy, $s$
\citep[\eg][\S14]{LandauLifshitz05}.
We can then define the density, $\rho = 1/H_p$,
and temperature, $T = H_s$,
where the subscripts on $H$ represent partial derivatives.
The linear perturbations to $\rho$ and $T$ are then given by
\begin{align}
  -\frac{\rho_1}{\rho_0^2} &= H_{ps}s_1 + H_{pp}p_1 \label{eq:FC_rho} \\
  T_1 &= H_{ss}s_1 + H_{sp}p_1, \label{eq:FC_T}
\end{align}
with similar expressions applying to the spatial variations of the background state:
\begin{align}
  -\frac{\bnabla\rho_0}{\rho_0^2} &= H_{ps}\bnabla s_0 + H_{pp}\bnabla p_0 \label{eq:back_rho} \\
  \bnabla T_0 &= H_{ss}\bnabla s_0 + H_{sp}\bnabla p_0. \label{eq:back_T}
\end{align}
(Note that $H_{ps} = H_{sp}$.)
The second derivatives
in \eqs(\ref{eq:FC_rho})--(\ref{eq:back_T})
can also be expressed in terms of the sound speed, $c$, and specific heats, $\cp$ and $\cv$,
as follows
\begin{align}
  H_{pp} = -\dfrac{1}{\rho_0^2c^2}, \qquad
  H_{ss} = \dfrac{T_0}{\cp}, \qquad
  \frac{H_{ps}^2}{H_{pp}H_{ss}} = 1 - \frac{\cp}{\cv}.
  \label{eq:coeffs}
\end{align}
Finally, the buoyancy frequency, $N$, is defined by the formula
$N^2 = -H_{ps}\dfrac{\dd p_0}{\dd z}\dfrac{\dd s_0}{\dd z}$.
We are concerned here with convective instability, which requires that $N^2 < 0$.
For notational convenience, we therefore define the parameter $\N = -N^2$
as a measure of the degree of superadiabaticity.
In what follows we will assume that $\N$ is positive throughout the domain, but we will make no assumption about its magnitude.
Note that the right-hand side of the momentum equation~(\ref{eq:FC_mom}) can be
written in many different forms, by using the relations (\ref{eq:FC_rho})--(\ref{eq:back_T})
and the hydrostatic condition $\bnabla p_0 = \rho_0\boldsymbol{g}$;
the form used here has been chosen for later convenience.

The linearised Boussinesq equations, for comparison, are
\begin{align}
  \frac{\partial\boldsymbol{u}}{\partial t} + 2\boldsymbol{\Omega}\times\boldsymbol{u}
  &= \frac{\rho_1}{\rho_0}\boldsymbol{g} - \frac{1}{\rho_0}\bnabla p_1
    + \boldsymbol{f}_1 \label{eq:Bous_mom} \\
  \bnabla\bcdot\boldsymbol{u} &= 0 \\
  \frac{\partial s_1}{\partial t} + \boldsymbol{u}\bcdot\bnabla s_0
  &= \frac{q_1}{T_0} \\
  -\frac{\rho_1}{\rho_0^2} &= H_{ps}s_1
  \label{eq:Bous_rho} \\
  T_1 &= H_{ss}s_1, \label{eq:Bous_T}
\end{align}
where $\rho_0$ is approximated as a constant, but $s_0$ retains its dependence on $z$.
In the Boussinesq approximation, the viscous force and the heating term take the form
\begin{align}
  \boldsymbol{f}_1 = \bnabla\bcdot(\nu\bnabla\boldsymbol{u})
  \qquad \mbox{and} \qquad
  \frac{q_1}{T_0} = \bnabla\bcdot(\kappa\bnabla s_1)
\end{align}
where $\nu$ is the kinematic viscosity and $\kappa$ is the thermal diffusivity.
The derivation of these Boussinesq equations relies upon
the following assumptions \citep[\eg][]{SpiegelVeronis60,Mihaljan62}:
\begin{enumerate}
  \item the domain height, $L$ say, is much smaller than the scale height of each thermodynamic variable; \label{height}
  \item perturbations to the thermodynamic variables are even smaller than the variations in their background values;
  \label{small}
  \item fluid motions are slow in comparison with the sound speed, $c$; \label{Mach}
  \item any timescale of the flow is much longer than the acoustic timescale, $L/c$. \label{slow}
\end{enumerate}
Often these assumptions are not stated explicitly, but are implicit in the scalings assumed for the various physical quantities.
Many derivations incorporate additional assumptions in order to simplify the Boussinesq equations still further,
for example by neglecting viscous heating \citep{Veronis62,Mihaljan62,GrayGiorgini76}.
In what follows we examine the onset of oscillatory convective instability in the Boussinesq and compressible systems,
which renders such additional assumptions unnecessary, because viscous heating
vanishes for linear perturbations to a hydrostatic state.
Moreover, conditions (\ref{small}) and (\ref{Mach}) are automatically satisfied in this case,
because near onset the perturbations to the hydrostatic background state are infinitesimally small.
Our aim is therefore to determine whether
the Boussinesq and fully compressible systems
have the same convective stability threshold
in the regime described by conditions (\ref{height}) and (\ref{slow}).

\subsection{The linear stability of the Boussinesq system}

We first derive the stability threshold for the Boussinesq equations, summarising the results
of \citet{Chandrasekhar53}. Suppose, for simplicity, that $\nu$, $\kappa$, and $\N$
are positive and constant throughout the domain.
We can then seek solutions of \eqs(\ref{eq:Bous_mom})--(\ref{eq:Bous_T})
in which the perturbations are
$\propto \exp({\ii\boldsymbol{k}\bcdot\boldsymbol{x}-\ii\omega t})$,
and thus obtain a dispersion relation
\begin{equation}
  \frac{\N\kh^2}{(\kappa k^2 - \ii\omega)} = (\nu k^2 - \ii\omega)k^2 + \frac{4\Omega^2k_z^2}{(\nu k^2 - \ii\omega)},
\end{equation}
where $k=|\boldsymbol{k}|$ and $\kh = \sqrt{k_x^2+k_y^2}$.
The onset of instability can be either direct (\ie\ $\omega=0$) or oscillatory (\ie\ $\omega^2>0$)
depending on parameter values.
In the former case, a perturbation with given $\boldsymbol{k}$
is convectively unstable if
\begin{equation}
  \N > \nu\kappa\left(\frac{k^6}{\kh^2}\right) + \frac{\kappa}{\nu}\left(\frac{4\Omega^2k_z^2}{\kh^2}\right).
  \label{eq:RB-direct}
\end{equation}
In the latter case, such a perturbation is unstable if
\begin{equation}
  \tfrac{1}{2}\N > \nu(\kappa+\nu)\left(\frac{k^6}{\kh^2}\right) + \frac{\nu}{\kappa+\nu}\left(\frac{4\Omega^2k_z^2}{\kh^2}\right),
  \label{eq:RB-over}
\end{equation}
and, at onset, it oscillates with a frequency given by
\begin{equation}
  \omega^2 = \frac{\kappa-\nu}{\kappa+\nu}\left(\frac{4\Omega^2 k_z^2}{k^2}\right) - \nu^2k^4.
\end{equation}
In the simplest case of Rayleigh--B\'enard convection between horizontal plates separated by distance $L$,
with fixed-temperature and stress-free boundary conditions,
the most unstable mode can be found by simply setting $k_z = \pi/L$
and finding the minimum unstable value of $\N$ over all $\kh$.
From an examination of \eqs(\ref{eq:RB-direct}) and (\ref{eq:RB-over}) it can be seen that
oscillatory instability is favoured in a rapidly rotating fluid with low viscosity,
and that in such cases the instability broadly resembles a growing inertial wave.
In the double asymptotic limit with $\nu \ll \kappa \ll \Omega L^2$ the unstable modes
are quasi-geostrophic ($\kh^2 \gg k_z^2$),
and the instability boundary is approximately described by
\begin{flalign}
  && \tfrac{1}{2}\N &\simeq \nu\kappa \kh^4 + \frac{\nu}{\kappa}\left(\frac{4\Omega^2k_z^2}{\kh^2}\right)& \label{eq:Bous} \\
  &\mbox{and}& \omega^2 &\simeq \frac{4\Omega^2k_z^2}{\kh^2}.& \label{eq:Bous2}
\end{flalign}
Minimising over $k_h$, we obtain the asymptotic scalings for the instability at onset:
\begin{align}
  \kh &\simeq \left(\sqrt{2}\Omega k_z/\kappa\right)^{1/3} \label{eq:scale_kh} \\
  \omega &\simeq \left(4\sqrt{2}\Omega^2\kappa k_z^2\right)^{1/3} \label{eq:scale_omega} \\
  \N &\simeq 6\left(2\Omega^2\kappa k_z^2\right)^{2/3}\nu/\kappa. \label{eq:scale_N2}
\end{align}

It is sometimes helpful to rewrite these results in terms of dimensionless quantities.
For Boussinesq convection, the standard dimensionless numbers are the Rayleigh, Taylor, and Prandtl numbers,
defined respectively as
\begin{equation}
  \Ray \equiv \frac{\N L^4}{\nu\kappa},
  \qquad
  \Tay \equiv \frac{4\Omega^2L^4}{\nu^2},
  \qquad
  \Pran \equiv \frac{\nu}{\kappa}.
\end{equation}
After putting $k_z=\pi/L$, \eq(\ref{eq:scale_N2}) can be rewritten as
\begin{equation}
  \Ray \simeq \tfrac{3}{2}(4\pi^2\Pran^2\Tay)^{2/3}
  \label{eq:Chand}
\end{equation}
\citep{Chandrasekhar53} and the double asymptotic limit in which this result applies can be expressed as
$1 \ll \Pran^{-1} \ll \Tay^{1/2}$.
\Eqs(\ref{eq:scale_kh}) and (\ref{eq:scale_omega})
imply that $\omega \simeq \sqrt{2}\kappa\kh^2$,
so the horizontal lengthscale of the instability is such that heat can diffuse between the upflows and downflows
on the same timescale as their oscillation period.
The instability can be understood physically as an inertial wave that is damped by viscosity
but amplified by thermal diffusion, which feeds heat into the upflows,
making them buoyant,
thereby extracting potential energy from the background stratification.

\subsection{The linear stability of the fully compressible system}
\label{sec:compress}

We now consider the stability properties of the fully compressible equations (\ref{eq:FC_mom})--(\ref{eq:FC_T}).
We recall that these equations are valid for an arbitrary equation of state, and an arbitrary hydrostatic background,
so in general their solutions will have complex, non-sinusoidal dependence on $z$.
Nevertheless we can still seek solutions with a given horizontal wavenumber $\kh$ and frequency $\omega$.
Our aim is to determine in what asymptotic limit (if any) the fully compressible equations
have the same instability threshold as the Boussinesq equations.
Assuming that such a limit does exist, we anticipate that the unstable modes will
follow the Boussinesq scaling laws (\ref{eq:scale_kh})--(\ref{eq:scale_N2}),
and in particular that $\kh \gg 1/L$.
In that case,
we can neglect vertical diffusion, and approximate the diffusive terms in equations
(\ref{eq:FC_mom}) and (\ref{eq:entropy}) as
\begin{align}
  \boldsymbol{f}_1 \simeq -\nu\kh^2\boldsymbol{u}
  \qquad \mbox{and} \qquad
  q_1 \simeq -\cp\kappa\kh^2T_1. \label{eq:FC_q}
\end{align}
(Note that we use the correct definition of $\kappa$ given by \citet{SpiegelVeronis60},
rather than that of \citet{Chandrasekhar61}.)
The validity of this approximation will be confirmed later, by solving the full set of linear equations numerically.
With this simplification, the linear equations (\ref{eq:FC_mom})--(\ref{eq:FC_T}) can be reduced to a pair of coupled ordinary differential equations for the perturbation quantities $p_1$ and $u_z$:
\begin{align}
  \left[\frac{\dd}{\dd z} + \frac{g}{c^2}\left(\frac{\frac{\cp}{\cv}\kappa\kh^2-\ii\omega}{\kappa\kh^2-\ii\omega}\right)\right]p_1
  &= \left[\frac{\N}{\kappa\kh^2-\ii\omega}-(\nu\kh^2-\ii\omega)\right]\rho_0u_z
  \label{eq:dpdz} \\
  \left[\frac{\dd}{\dd z} + \frac{\N}{g}\left(\frac{\ii\omega}{\kappa\kh^2-\ii\omega}\right)\right]\rho_0u_z
  &= \left[\frac{\ii\omega}{c^2}\left(\frac{\frac{\cp}{\cv}\kappa\kh^2-\ii\omega}{\kappa\kh^2-\ii\omega}\right) - \frac{(\nu\kh^2-\ii\omega)\kh^2}{4\Omega^2+(\nu\kh^2-\ii\omega)^2}\right]p_1.
  \label{eq:dudz}
\end{align}

In order to make contact with the Boussinesq results, we must now impose the restrictions built into the Boussinesq approximation.
First, we take the limit in which the height of the domain, $L$, is much smaller than any lengthscale on which the
background state varies.
This corresponds to condition~(\ref{height}) given earlier.
In this limit, and on the assumption that the scaling $\omega \sim \kappa\kh^2$ is still valid,
the left-hand sides of \eqs(\ref{eq:dpdz}) and (\ref{eq:dudz}) are dominated by the derivative terms,
and the parenthetical terms on the right-hand sides can be approximated as constant.
Under condition (i), it is reasonable to expect growing modes near onset to have a sinusoidal dependence on $z$, which means that we can replace $\frac{\dd}{\dd z} \to \ii k_z$, thus obtaining
an approximate dispersion relation
\begin{equation}
  0 \simeq k_z^2
  + \left[\frac{\N}{\kappa\kh^2-\ii\omega}-(\nu\kh^2-\ii\omega)\right]
    \left[\frac{\ii\omega}{c^2}\left(\frac{\frac{\cp}{\cv}\kappa\kh^2-\ii\omega}{\kappa\kh^2-\ii\omega}\right) - \frac{(\nu\kh^2-\ii\omega)\kh^2}{4\Omega^2+(\nu\kh^2-\ii\omega)^2}\right].
  \label{eq:dispersion}
\end{equation}

Next, we assume that the oscillation frequency of the instability is much smaller than
the resonant acoustic frequency (\ie\ $\omega \ll c/L$),
in accordance with condition~(\ref{slow}).
Provided that the scalings (\ref{eq:scale_kh})--(\ref{eq:scale_N2})
are still correct in order of magnitude,
the real and imaginary parts of the dispersion relation (\ref{eq:dispersion})
imply that
\begin{flalign}
  && \tfrac{1}{2}\N &\simeq \nu\kappa \kh^4 + \frac{4\Omega^2k_z^2}{\kh^2}\left[\frac{\nu}{\kappa}
  + \left(\frac{\cp}{\cv}-1\right)\left(\frac{2\Omega^2}{c^2\kh^2}\right)\right]& \label{eq:N2} \\
  &\mbox{and}& \omega^2 &\simeq \frac{4\Omega^2k_z^2}{\kh^2}& \label{eq:freq}
\end{flalign}
at the onset of instability.
We note that the imaginary part of \eq(\ref{eq:dispersion}) is smaller than the real part
by a factor of $\Pran = \nu/\kappa \ll 1$.  Therefore in order to obtain \eqs(\ref{eq:N2})--(\ref{eq:freq})
it is not sufficient to consider only the leading-order terms in \eq(\ref{eq:dispersion}).

Comparing \eqs(\ref{eq:Bous}) and (\ref{eq:N2}),
we see that the Boussinesq results are valid only if
$(\nu/\kappa)^{1/2}c \gg \Omega/\kh \sim \omega L$,
which is much more restrictive than the condition $c \gg \omega L$ assumed above.
Conditions (\ref{height})--(\ref{slow}) are therefore \emph{not} sufficient
to guarantee the validity of the Boussinesq approximation.
The additional term in \eq(\ref{eq:N2}), compared with \eq(\ref{eq:Bous}), renders the fully compressible system subject to conditions~(\ref{height})--(\ref{slow}) more stable than the equivalent Boussinesq system, because each wavenumber becomes marginally stable at a larger value of $\N$.

The discrepancy between the true instability threshold and that predicted by the Boussinesq equations
suggests that some effect of compressibility remains significant even under conditions (\ref{height})--(\ref{slow}).
In the compressible system, \eqs(\ref{eq:entropy}) and (\ref{eq:FC_q})
describe the advection of \emph{entropy} and the diffusion of \emph{temperature}.
If we rewrite \eq(\ref{eq:entropy}) in terms of the density and pressure perturbations,
then we obtain
\begin{align}
  \frac{\rho_1}{\rho_0} = -\frac{\N u_z/g}{\kappa\kh^2-\ii\omega}
    + \left(\frac{\tfrac{\cp}{\cv}\kappa\kh^2-\ii\omega}{\kappa\kh^2-\ii\omega}\right)\frac{p_1}{\rho_0c^2}.
\end{align}
The term in parentheses is complex, and its imaginary part quantifies the density perturbations arising from
pressure perturbations that are \emph{out of phase}.  This is the source of the extra term in \eq(\ref{eq:N2})
compared with \eq(\ref{eq:Bous}).
In the Boussinesq equations, the density, temperature and entropy
perturbations are all proportional to one another (\eqs\ref{eq:Bous_rho} and \ref{eq:Bous_T}),
so this effect is absent.
We note that this discrepancy does not arise in direct (\ie\ non-oscillatory) convection,
for which all perturbations are necessarily in phase with one another,
so our findings apply only to oscillatory instabilities.

\subsection{A specific example: The case of an ideal gas}

To more precisely illustrate the discrepancy between the Boussinesq and compressible results,
we now consider the particular case of an ideal gas.
With this simplification, the compressible system can be defined in terms of six dimensionless parameters.
It is convenient to choose three of these to be $\Ray$, $\Tay$ and $\Pran$, as in the Boussinesq system.
For the remaining three, we will choose
the ratio of specific heats, $\cp/\cv$,
the ratio of the temperature gradient to the adiabatic temperature gradient,
$\dfrac{\nabla}{\nabla_{\rm ad}} = 1 + \dfrac{H_{ss}}{H_{sp}}\dfrac{\dd s_0}{\dd p_0}$,
and the ratio of the domain height to the temperature scale height, $\theta = L/h_T$,
where $h_T \equiv -(\dd\ln T_0/\dd z)^{-1}$.
Note that, in a domain of finite height, all of these parameters except $\cp/\cv$ will generally depend on $z$.
However, if the domain is sufficiently shallow, as we will shortly assume, then they may be approximated as constant.
For simplicity we will fix $\cp/\cv=5/3$ and $\nabla/\nabla_{\rm ad} = 2$,
so that $\theta$ is the only additional variable in our compressible system.
We anticipate that the system will become Boussinesq in the limit $\theta\to0$,
but we need to determine exactly how small $\theta$ must be for the Boussinesq results to hold.
Given our choices for the other two parameters,
the scale heights of pressure and density are
$h_p \equiv -(\dd\ln p_0/\dd z)^{-1} = \frac{4}{5}L/\theta$ and
$h_\rho \equiv -(\dd\ln\rho_0/\dd z)^{-1} = 4L/\theta$, and we have
\begin{equation}
  \frac{\N}{c^2} = \frac{3}{8}(\theta/L)^2. \label{eq:theta}
\end{equation}
Because all of the thermodynamic scale heights are of order $L/\theta$,
condition~(\ref{height}) is simply $\theta \ll 1$.
Condition~(\ref{slow}), which requires that $\omega \ll c/L$,
turns out to be more strict.  Combining \eq(\ref{eq:theta}) with (\ref{eq:scale_omega}) and (\ref{eq:scale_N2}),
this condition becomes $\theta \ll (8\,\Pran)^{1/2}$.
But even when this condition is satisfied, the Boussinesq results may not be valid.
In fact, the marginal stability criterion (\ref{eq:N2}) then becomes, in terms of our chosen dimensionless parameters,
\begin{equation}
  \Ray \simeq 2\kh^4 + \Pran^2\Tay\frac{2\pi^2}{\kh^2} + \frac{\Pran^2\Tay^2}{\Ray}\frac{\pi^2\theta^2}{4\kh^4}.
  \label{eq:crit}
\end{equation}
Here $k_h$ and $k_z$ are measured in units of $1/L$,
and we have put $k_z = \pi$, assuming stress-free, fixed-temperature boundary conditions.
This expression can be solved for $\Ray$, and the critical Rayleigh number is then found by
minimising over all $\kh$.  The result is
\begin{flalign}
  && \Ray &\simeq (1+X^{-1})(2\pi^2X\Pran^2\Tay)^{2/3},&
  \label{eq:true} \\
  &\mbox{where}& X &= \frac{1}{2} + \sqrt{\frac{9}{4} + \frac{\theta^2}{2\pi^2\Pran^2}}.
\end{flalign}
We recover the Boussinesq result, which corresponds to $X=2$, only if $\theta \ll \Pran$.
Under astrophysical conditions, this is a rather extreme restriction.  In the solar interior, for example, the Prandtl number
is of order $\Pran \simeq 10^{-6}$, and the temperature scale height is of order $10^5$\,km.
So our result implies that the Boussinesq approximation is only valid on vertical scales smaller than 1\,km!
Of course, convection in the solar interior is well developed, and so the linear stability analysis presented here
is not directly applicable.  However, there is no obvious reason why the Boussinesq equations should be any more
valid in the strongly nonlinear regime than they are in the linear one.
Moreover, numerical simulations of the Sun (and, indeed, other convective systems)
are never performed very far above the convective threshold,
owing to computational constraints, and so our results may well be directly applicable to those simulations.

\begin{figure}
  \centering%
  \includegraphics[height=0.465\textwidth]{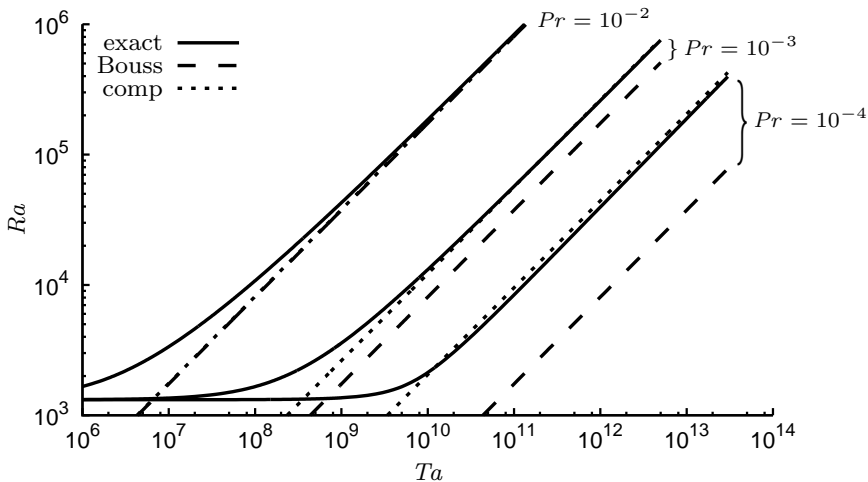}%
  \caption{The critical Rayleigh number for onset of oscillatory convection as a function of Taylor number
  in a compressible fluid (solid curves) for three different Prandtl numbers.
  The dashed and dotted lines show the asymptotic scalings
  given by \eqs(\ref{eq:Chand}) and (\ref{eq:true}) respectively.}%
  \label{fig:Bushby}%
\end{figure}

In order to confirm the validity of these results, we have used a linear eigensolver to compute the critical
Rayleigh number as a function of Taylor number in an ideal gas with $\theta=0.02$
for three different values of $\Pran$, with other parameters
as given above.
The solver, which uses the algorithm originally developed by \citet{Gough-etal76},
solves the exact linearised equations for an ideal gas with constant dynamic viscosity and thermal conductivity.
We consider the simplest case of stress-free, fixed-temperature boundary conditions,
so that the exact stability threshold can be directly compared with the asymptotic scalings
obtained from the Boussinesq (\ref{eq:Chand}) and fully compressible (\ref{eq:true}) equations.
This comparison is shown in Figure~\ref{fig:Bushby}.
In the rapidly rotating regime, with $\Tay \gg 1/\Pran^2$, the stability threshold closely matches that predicted by \eq(\ref{eq:true}) in each case.  As the Prandtl number is decreased, the Boussinesq prediction becomes less
accurate, and always underestimates the true stability threshold.

\section{Other sound-proof models}

We have shown that the Boussinesq equations
generally fail to predict the true onset of oscillatory convection,
unless the height of the domain
is \emph{much} smaller than
a typical thermodynamic scale height,
by a factor of less than $\Pran \ll 1$.
However, there are alternatives to the Boussinesq equations that purport to be valid even on scales
larger than a typical scale height.
We might then wonder whether these other ``sound-proof'' approximations, such as the anelastic
and \PI\ approximations, more accurately predict the onset of oscillatory convection.
In fact, some implementations of the anelastic approximation certainly do not perform better:
the anelastic model of \citet{Drew-etal95} produces spurious convective instability in some cases with a negative Rayleigh number.
\citet{Calkins-etal15}, using a similar anelastic model,
showed that the discrepancy between the anelastic and fully compressible equations
becomes increasingly serious at smaller Prandtl numbers, which is reminiscent of the results shown in \fig\ref{fig:Bushby}.
Before proceeding, it is worth reviewing some important points about the different sets of sound-proof
equations, and the origins of their multiplicity.

As originally shown by \citet{EckartFerris56}, the linearized, fully compressible equations
(\ref{eq:FC_mom})--(\ref{eq:FC_T})
have an energy principle
\begin{align}
  \frac{\partial E}{\partial t}
  &+ \bnabla\bcdot(p_1\boldsymbol{u})
  = \rho_0\boldsymbol{u}\bcdot\boldsymbol{f}_1 + \left(T_1-\frac{\dd T_0}{\dd s_0}s_1\right)\frac{\rho_0q_1}{T_0},
  \label{eq:energy}
\end{align}
where $E$ is the ``available'' or ``external'' energy
\begin{equation}
  E = \frac{1}{2}\rho_0\left(u^2 + \frac{N^2s_1^2}{(\dd s_0/\dd z)^2} + \frac{p_1^2}{\rho_0^2c^2}\right).
\end{equation}
The three contributions to $E$ are generally referred to as ``kinetic'', ``thermobaric'', and ``elastic'', respectively.

It can be shown that the Boussinesq equations also satisfy \eq(\ref{eq:energy}),
provided that the available energy is redefined as
\begin{equation}
  E = \frac{1}{2}\rho_0\left(u^2 + \frac{N^2s_1^2}{(\dd s_0/\dd z)^2}\right),
  \label{eq:anelastic}
\end{equation}
indicating that these equations do not support elastic motions (\ie\ sound waves).
However, the Boussinesq equations assume that the background density $\rho_0$ is constant.
\citet{OguraPhillips62} sought to extend the Boussinesq equations to more general background states,
and they recognised the importance of preserving \eq(\ref{eq:energy}), with $E$ defined as in \eq(\ref{eq:anelastic}).  They referred to this as the ``anelastic'' approximation.
Subsequently, however, the meaning of this term has become distorted, and most sets of equations in use today
that are referred to as anelastic do not actually satisfy \eq(\ref{eq:energy}) for \emph{any} definition of $E$
\citep[\eg\ see review by][]{Brown-etal12}.
Instead, the term ``anelastic'' is now used exclusively to describe sets of equations that include the velocity constraint
\begin{equation}
  \bnabla\bcdot(\rho_0\boldsymbol{u}) = 0 \label{eq:anelastic2}
\end{equation}
\citep[\eg][]{BraginskyRoberts07}.
Rather than appealing to energy conservation, most derivations of the anelastic equations have relied
entirely on formal asymptotics,
writing each term in the equations as a truncated expansion in powers of one or more hypothetically small parameters.
This approach has some potential pitfalls.
First, the resulting set of equations may not be
unique: for example, by expressing the original, exact equations in slightly different forms,
different sets of anelastic equations can be obtained as first-order truncations
under the same asymptotic conditions \citep[\eg][]{Berkoff-etal10}.
Second, as demonstrated by the unphysical ``negative Rayleigh number'' convection
seen in some implementations
of the anelastic equations,
there is no guarantee that equations obtained from
a truncated asymptotic expansion
will exhibit sensible behaviour.
In some problems, important information is contained in the higher-order terms,
though this may not be obvious at first sight.
For instance, we have shown
in section~\ref{sec:compress}
that the phase shift between pressure and density is crucial in oscillatory convection,
and that this information is lost if we consider only the leading-order terms in the equations.

We therefore suggest an alternative approach in deriving sound-proof equations,
which is to take the energy principle idea
of \citet{OguraPhillips62}
and follow it to its logical conclusion.
Specifically, we ask ``what is the most general set of linear equations that is consistent with
\eqs(\ref{eq:energy}) and (\ref{eq:anelastic})?''
In order for our equations to be physically meaningful, we will require that they also obey conservation
of mass and entropy, \ie\ our set of equations must include (\ref{eq:mass}) and (\ref{eq:entropy}),
and we require that $\rho_1$ and $T_1$ are linearly related to $s_1$ and $p_1$, with coefficients
that depend only on the background state locally.
(Note, however, that the background state must still satisfy the exact relations
(\ref{eq:back_rho}) and (\ref{eq:back_T}).)
Finally, we require that the momentum equation
includes a pressure gradient force of the form $-(1/\rho_0)\bnabla p_1$,
and a buoyancy force that acts only in the vertical direction.
The form of this buoyancy force we leave unspecified,
though it must be a linear function of the thermodynamic perturbations.
So we consider the equations
\begin{align}
  \frac{\partial\boldsymbol{u}}{\partial t} + 2\boldsymbol{\Omega}\times\boldsymbol{u}
  &= (A_pp_1+A_\rho\rho_1+A_ss_1+A_TT_1)\boldsymbol{e}_z - \frac{1}{\rho_0}\bnabla p_1
    + \boldsymbol{f}_1 \\
  \rho_1 &= B_ss_1 + B_pp_1 \\
  T_1 &= C_ss_1 + C_pp_1,
\end{align}
and we ask what conditions on the coefficients
$A_X, B_X, C_X$
are imposed by equations (\ref{eq:mass}), (\ref{eq:entropy}),
(\ref{eq:energy}) and (\ref{eq:anelastic}).
Perhaps surprisingly,
after some straightforward (but lengthy) algebra, it can be shown that these restrictions
remove almost all freedom in the choice of coefficients!
The most general set of equations permitted is
\begin{align}
  \frac{\partial\boldsymbol{u}}{\partial t} + 2\boldsymbol{\Omega}\times\boldsymbol{u}
  &= - s_1\bnabla T_0 + T_1\bnabla s_0 - \bnabla(p_1/\rho_0)
    + \boldsymbol{f}_1 \label{eq:PI_mom} \\
  \frac{\partial\rho_1}{\partial t} + \bnabla\bcdot(\rho_0\boldsymbol{u}) &= 0
  \label{eq:PI_mass} \\
  \frac{\partial s_1}{\partial t} + \boldsymbol{u}\bcdot\bnabla s_0
  &= \frac{q_1}{T_0} \label{eq:PI_s} \\
  -\frac{\rho_1}{\rho_0^2} &= \alpha H_{ps} s_1 \label{eq:PI_rho} \\
  T_1 &= H_{ss}s_1 + \alpha H_{ps} p_1, \label{eq:PI_T}
\end{align}
where the only remaining free parameter,
which we have chosen to call $\alpha$, is an arbitrary function of the background state.
Remarkably, the momentum equation~(\ref{eq:PI_mom}) necessarily takes exactly the form of \eq(\ref{eq:FC_mom}),
so the only differences between the fully compressible equations and these sound-proof equations
occur in the linearised equations of state, (\ref{eq:PI_rho}) and (\ref{eq:PI_T}). To the best of our knowledge,
\eqs(\ref{eq:PI_mom})--(\ref{eq:PI_T}) have never before been written down in exactly this form.
They are similar to those of \citet{CotterHolm14},
but more general because they include a general equation of state
and non-adiabatic effects,
and they also provide a unique and consistent prescription for the temperature perturbation $T_1$.
We can determine the onset of oscillatory convection in these equations by following the same procedure
that led to \eq(\ref{eq:N2}).  Under the ``Boussinesq'' conditions (i)-(iv),
and approximating the diffusive terms using \eq(\ref{eq:FC_q}),
we find that marginal stability is achieved when
\begin{align}
  \tfrac{1}{2}\N &\simeq
  \nu\kappa\kh^4
  + \frac{4\Omega^2k_z^2}{\kh^2}\left[\frac{\nu}{\kappa}
  + \alpha^2\left(\frac{\cp}{\cv}-1\right)\left(\frac{2\Omega^2}{c^2\kh^2}\right)\right].
  \label{eq:N2_alpha}
\end{align}
For any non-zero value of $\alpha$ this implies that
the onset of oscillatory convection in
this system (at least under the standard Boussinesq conditions) occurs at a larger value of $\N$, and therefore at a higher Rayleigh number, than in the Boussinesq system.

\section{Discussion}
\label{sec:discussion}

Although we have referred to \eqs(\ref{eq:PI_mom})--(\ref{eq:PI_T}) as ``sound-proof'',
we have not explicitly shown that acoustic oscillations are absent from these equations.
That this is the case can be demonstrated, for example, by deriving their exact dispersion relation
for perturbations about some idealised background state.  However, it can also be deduced simply from \eq(\ref{eq:PI_rho}),
which describes how the volume of a displaced parcel of fluid instantaneously adjusts to a value determined
by the entropy inside the parcel and the local properties of the background state.
By neglecting the effect of pressure perturbations on this adjustment, we remove the dynamical degree of
freedom that permits the parcel to oscillate acoustically.  A mathematical proof of this statement, which implicitly
assumes the conservation of mass and entropy, has been given by \citet{Durran08}.

The direct relation between $\rho_1$ and $s_1$ imposes a constraint on the velocity field,
which can be deduced from \eqs(\ref{eq:PI_mass})--(\ref{eq:PI_rho}):
\begin{equation}
  \frac{1}{\alpha H_{ps}\rho_0^2}\bnabla\bcdot(\rho_0\boldsymbol{u}) + \boldsymbol{u}\bcdot\bnabla s_0
  = \frac{q_1}{T_0}. \label{eq:constraint}
\end{equation}
Although $\alpha$ can be taken as any function of height,
there are two significant special cases, which correspond to the particular choices $\alpha=0$ and $\alpha=1$.
When $\alpha=0$, \eq(\ref{eq:constraint}) reduces to the anelastic velocity constraint (\ref{eq:anelastic2})
and, in fact,
our equations become almost identical to the
particular version of the anelastic equations derived by \citet{Lantz92} and by \citet{BraginskyRoberts95},
which for brevity we will call the LBR equations.
However, an important difference is that in our sound-proof system, with $\alpha=0$,
\eq(\ref{eq:PI_rho}) states that $\rho_1=0$,
whereas in the LBR equations $\rho_1$ is given as a function of the other thermodynamic
perturbations.  Formally this implies that mass is not conserved in the LBR equations,
whereas mass conservation is built in to our linearised system.
Nevertheless, because $\rho_1$ does not appear explicitly in \eqs(\ref{eq:PI_mom}) and (\ref{eq:PI_T}),
our equations are in fact mathematically equivalent to the LBR equations,
and the only difference is conceptual.
\citet{Lantz92} argued that these equations are the most natural generalisation of the Boussinesq
equations to a density-stratified background and, indeed, we see that \eq(\ref{eq:N2_alpha})
becomes identical to its Boussinesq counterpart (\ref{eq:Bous}) in the case where $\alpha=0$ (this means that the curves in \fig\ref{fig:Bushby} labelled as ``Boussinesq'' also correspond to the $\alpha=0$ case of our sound-proof equations).

A significant practical advantage of the LBR equations over other sound-proof approximations
is that they can be solved without ever having to
explicitly calculate the pressure perturbation $p_1$, provided that perturbations to the heat flux are calculated
from the gradient of $s_1$ rather than $T_1$ \citep{Lantz92,BraginskyRoberts95}.
In previous derivations of the LBR equations, this approximation has been justified by appealing to
``turbulent diffusion'' of entropy.  However, setting $\alpha=0$ implies a direct
relation between $T_1$ and $s_1$ in \eq(\ref{eq:PI_T}).  So in our version of the anelastic system there is no need to approximate the
heat flux; instead we have an approximate equation of state (\ref{eq:PI_T}) whose form is dictated
by the energy principle (\ref{eq:energy}).

Whilst the choice $\alpha=0$ reduces \eqs(\ref{eq:PI_mom})--(\ref{eq:PI_T}) to the LBR anelastic equations,
\eq(\ref{eq:N2_alpha}) suggests that the ``best'' sound-proof model is actually given by $\alpha = 1$, because in that case we exactly recover the corresponding compressible result (\eq\ref{eq:N2}) from \eq(\ref{eq:N2_alpha}).
(The choice $\alpha = -1$ can be discounted on physical grounds, because it would imply a positive correlation between $\rho_1$ and $s_1$.)
If we set $\alpha=1$ then \eqs(\ref{eq:PI_mom})--(\ref{eq:PI_T}) are equivalent to the linearisation of the ``thermodynamically consistent'' version of the \PI\ equations obtained by \citet{KleinPauluis12} (see also the ``generalized \PI'' equations of \citet{Vasil-etal13}). In fact, the results of \citet{KleinPauluis12} suggest how the method that we have proposed for standardising sound-proof models can be extended into the nonlinear regime. This is an essential step towards modelling systems that are well above the onset of convection, such as the interiors of stars, but is beyond the scope of the current paper.

A very similar set of \PI\ equations was recently studied numerically by \citet{Lecoanet-etal14}.
An important difference, however, is in the definition of the temperature perturbation $T_1$.
We believe that many of the discrepancies \citet{Lecoanet-etal14} found between their solutions of the \PI\
equations and the fully compressible equations result from the incorrect definition of $T_1$ that they used,
but further work will be required to confirm this.  Having said that, not all of the issues with the \PI\ equations identified by \citet{Lecoanet-etal14} can be solved simply by redefining $T_1$.  In particular, the velocity constraint (\ref{eq:constraint})
becomes ill-posed for horizontally-invariant perturbations in a fluid with impenetrable, thermally conducting boundaries.
This difficulty does not arise in the stability analysis presented here, because convective motions necessarily have a finite
horizontal lengthscale. In general, the \PI\ equations seem best suited to describing fluid motions on small horizontal scales, and are less accurate for motions on large horizontal scales \citep[see \eg][]{Durran08}.
One possible resolution is to allow the background state to be time dependent \citep[\eg][]{ONeillKlein14},
but this is beyond the scope of the analysis presented here.

We would like to thank Professor Rupert Klein, as well as the anonymous referees, for their helpful comments and suggestions.

\end{document}